# Low-cost high-efficiency solar steam generation by wick material with graphite micro/nano particles

Guilong Peng[1,2, #], Hongru Ding[1,2, #], S.W. Sharshir[1,2,3], Dengke Ma[1, 2], Lirong Wu[1, 2], Jianfeng Zang[4,5], Huan Liu[4], Wei Yu[6], Huaqing Xie[6], Nuo Yang[1, 2, *]

1 State Key Laboratory of Coal Combustion, Huazhong University of Science and Technology, Wuhan 430074, China

2 Nano Interface Center for Energy (NICE), School of Energy and Power Engineering, Huazhong University of Science and Technology, Wuhan 430074, China

3 Mechanical Engineering Department, Faculty of Engineering, Kafrelsheikh University, Kafrelsheikh, Egypt.

4 School of Optical and Electronic Information, Huazhong University of Science and Technology, Wuhan 430074, China

5 Innovation Institute Huazhong University of Science and Technology, Wuhan 430074, China

6 School of Environment and Materials Engineering, College of Engineering, Shanghai Polytechnic University, Shanghai 201209, China

[#]Guilong Peng and Hongru Ding contribute equally on this work.

*Corresponding email: nuo@hust.edu.cn

**Abstract**

Generating water steam by solar energy is a significant process for many fields. In this paper, a low-cost high-efficiency wick type steam generator is proposed. It's based on the heat localization and thin-film evaporation. The measurements show that the energy efficiency is 84 % at 1 kw/m$^2$. Besides, the dependence of efficiency on particle concentration and size are discussed. The optimal particle concentration is found at 60 g/m$^2$, and a smaller particle size gives higher efficiency. The experimental results agree well with the theoretical prediction based on thin-film evaporation theory. Our study offers a new in-depth understanding of low-cost high-efficiency solar steam generation.

1. **Introduction**

Solar energy is abundant, renewable and eco-friendly, hence high efficiency solar energy harvesting technology becomes one of the most popular research topic. Solar steam generation technology, as one of the ways to utilize solar energy, has a lot of applications such as desalination [1-3], power generation [4-6], water purification [7], oil recovery [8, 9] and so on.

During the past decades, many works have been devoted to harvesting the solar energy for steam generation, for example, designing a high efficiency solar still [10-21]. Various types of solar still like the stepped solar still [18] and wick type solar still [19, 20] have been developed. However, the energy efficiency of those traditional solar stills are normally below 50% due to the large heat loss. Recently, a new effective method named heat localization was proposed to enhance the evaporation efficiency [22-29]. It's based on the thermal insulation between the evaporation region and the bulk water. The evaporation efficiency reaches up to 67% under 1 kw/m$^2$ by using a double layer structure, which is constructed by expanded graphite and carbon foam [23]. And the efficiency increases to 80% when the double layer structure is constructed by graphene oxide and polystyrene [24]. By using an aluminum nanoparticles coated nanoporous $Al_2O_3$ film, the efficiency will be 91% under 6 kw/m$^2$ insolation [29]. However, due to the complex design, the material cost in these researches is too high for large-scale applications.

Meanwhile, a promising strategy to enhance evaporation is utilizing the thin-film evaporation [30-35]. When liquid wets a solid wall, the extended meniscuses are typically divided into three regions: adsorbed region (I), thin-film region (II) and intrinsic meniscus region (III). The strongest evaporation happens at the surface of region II due to the very low thermal resistance across the liquid film. Therefore, the heat loss to the surrounding water is very low and the energy efficiency will increase. Micro [31] or nano [32] scale surface modifications and adding nanoparticles [33] or microparticles [34] in water are ways to increase the thin film region for higher energy efficiency. Besides, the efficiency can also be increased by improving the wettability of the solid wall [35].

In this paper, a low-cost high-efficiency wick type steam generator was proposed. It combines the heat localization and thin-film evaporation. Firstly, we studied the evaporation of bare water and steam generator with wick to show the effect of heat localization. Then, the effect of thin-film evaporation was studied quantitatively: we measured the evaporation rate of evaporation with micro/nano particles on wick and took the change of absorptivity into account. We also investigated the optimal concentration and size effect of particles, followed by the theoretical analysis of thin-film evaporation. In addition, the surfactant was used to further improve the efficiency.

## 2. Work principle and Experiment setup

The schematic diagram of evaporation from wick is shown in Fig 1a. Water is transported from the bottom to the top of the structure via the side wick by capillary effect, then heated by solar energy and evaporates. The optical microscopic image of wetted wick is shown in Fig 1b. The adiabatic foam can prevent the heat transfer from the top surface of the wick to the bulk water. Compared with the porous foam [23], which has relative high thermal conductivity when full filled with water, the waterproof type adiabatic material has better performance. The schematic diagram of the measuring setup is illustrated in Fig S1 at Supplemental Information.

To further increase the evaporation rate, graphite micro/nano particles were sprinkled and smeared uniformly on the top surface to create more thin-film region. Meanwhile, the high solar absorption and thermal conductivity of graphite particles [17, 36-38] can also improve the performance of the generator. The scanning electron microscope (SEM) images of graphite particles are shown in Fig S2. The optical microscopic image of wetted wick with graphite particles on it is presented in Fig 1c. It should be noted that the particles can be recycled easily by filtering as shown in Fig S3. The retail price of graphite particles is around $10/kg, which is very cheap compared to nano-materials.

The schematic diagram of the microscopic structure on the evaporation surface is shown in Fig 1d. The particles on the fiber form a porous structure whose surface is filled with meniscuses. As illustrated in the inset of Fig 1d, meniscus is constructed by three regions: (I) adsorbed or non-evaporation region, where water is adsorbed on the graphite due to the high disjoining pressure; (II) thin-film or transition region where effects of long-range molecular forces are felt; (III) intrinsic meniscus region, where the thickness of water layer increases very fast [39]. In the adsorbed region, water sticks to the graphite tightly and no mass/heat transfer occurs. Whereas in the thin-film region, the disjoin pressure is weak, while the thickness of water layer is still thin enough to assure a low thermal resistance. Therefore, the most heat current runs through the thin-film region, hence the fast evaporation and low heat loss.

Here, the black linen cloth and expanded polyethylene (EPE) were chosen as the wick material and adiabatic material, respectively (shown in Fig S4). However, some other low-cost daily materials can also be chosen as the wick material, such as cotton, tissue, paper and so on, as long as the capillary effect is strong enough to compensate the evaporation loss. The capillary ability of the wick material can be described by the rate of moisture regain, which is 12.5% for linen [40] and 8.5% for cotton [41]. The retail price of the linen cloth is around $5/m$^2$. On the other side, the adiabatic material should meet the demand of waterproof and floating on the water. The EPE is low-cost (retail price, ~$10/m$^3$), recyclable, anticorrosion and nontoxic. It's safer compared with other common adiabatic materials, like rubber insulation cotton and expanded polystyrene. The characteristics of the linen and EPE are listed in Table S1.

## 3. Results and discussion

The effects of the linen, EPE and graphite particles on the evaporation rate are shown in Fig 2a. The bulk bare water in a cup is around 60 g in mass and 5 cm in depth. The evaporation rate of the bare water is very low, and the accumulated mass change is less than 0.16 kg/m$^2$ in the first 30 minutes. The natural evaporation rate of the steam generator is measured at around 0.11 kg/(m$^2$·h) at room environment (temperature 24 ℃ and relative humidity 40%). All the values shown here has subtracted the natural evaporation rate already.

When using our steam generator with the wick (black linen) and EPE (W&E), the evaporated water reached to 0.39 kg/m$^2$ in the first half an hour, which is 2.5 times that of the bare water. When it reaches equilibrium state, the hourly evaporation rate is higher than the first non-equilibrium process and the value is 0.82 kg/(m²·h) for W&E. Moreover, we measured the system with graphite micro/nano particles. The results show that graphite particles can further enhance the evaporation. The accumulated mass change is 0.55 kg/m² in the first half an hour. When it reaches equilibrium state, the hourly evaporation is 1.15 kg/(m²·h) for generator with wick, EPE and graphite particles(W&E&G).

The relationship between the evaporation rate and energy efficiency is described as:

$$\eta = \frac{\Delta m \cdot h_{LV}}{Q} \tag{1}$$

where $\eta$ is the energy efficiency, $\Delta m$ is the evaporation rate, $h_{LV}$ is the total enthalpy of phase change, containing latent heat and sensible heat, $Q$ is the incident solar energy.

The instantaneous energy efficiency of evaporation is shown in Fig 2b. As we can see, the efficiency of evaporation from the bare water is continuously increasing during the measurement, which reaches to around 30% after half an hour. On the contrary, the maximum efficiency is reached in only 7 minutes when W&E or W&E&G are used. The energy efficiency is 56% and 79% for W&E and W&E&G, respectively. Obviously, the utilization of wick, EPE and graphite particles is very effective for enhancing the energy efficiency

To understand the high performance of evaporation from wick material, the absorptivity of different materials is measured firstly. As shown in Fig 2c, the absorptivity of water is around 10% at 1 cm thickness. But it reaches up to 83% for wetted wick, and further increases to 95% when the graphite particles are added on the wick material. Obviously, the higher absorptivity results in higher energy efficiency, but compared with the 23% increase on efficiency after added the graphite particles, the 12% increase on absorptivity is much lesser. It means that the better heat transfer performance which is resulted from the thin-film evaporation can enhance the efficiency by at least 11%.

Furthermore, the temperature is also measured to illustrate the heat localization performance of the proposed system. The surface temperature of the bare water keeps increasing during the measurement, which reaches to 34 ℃ after 30 minutes as shown in Fig 2c. The long non-equilibrium time indicates that a considerable proportion of the absorbed energy is used to increases the bulk water temperature instead of being used for evaporation, hence the low energy efficiency. On the other hand, the surface temperatures of W&E and

W&E&G reach up to around 44.5 °C and 48.5 °C in only 7 minutes, respectively, which is due to the significantly smaller water amount to be heat around the wick material. Meanwhile, the temperature of the bulk water under the EPE increases only 1 °C after 30 minutes, which means almost all the heat is localized at the evaporation surface and result in the high surface temperature and fast evaporation.

To investigate the best performance of our system, the effect of particle concentration on the efficiency was firstly studied. Although the absorptivity of different concentration is the same as shown in Fig S5, the efficiencies are different. As we can see in Fig 3a, the optimal concentration is found at 60 g/m$^2$, the efficiency will decrease slightly for less or more particles. The decrease may due to the insufficient thin-film region at low concentration and the impeded water supply by thick graphite layer at high concentration. Interestingly, when the concentration increases from 20 g/m$^2$ to 60 g/m$^2$, the efficiency only increases by 5%, but when the concentration increases from 0 to 20 g/m$^2$, the efficiency increases by around 18%. This result implies that the graphite particles play a very important role even when it is sparse.

Another important parameter, the particle size, is also considered. As shown in Fig. 3b, the evaporation rate is inversely proportional to the particle size, due to less particle-water interface and thin-film regions exist for larger particle size. When the size increases to 25 μm, the efficiency decreases by around 6% compared with that of 1.3 μm. The observed size effect is consistent with the theoretical analysis of thin-film evaporation as follows.

According to the thin-film evaporation theory [39,42-44]. The evaporation rate, ΔM, is given by (derivations shown in SI Note III):

$$\Delta M = \frac{C \cdot \alpha(d)}{d \cdot \delta(d)^2} \tag{2}$$

where $C$ is a constant coefficient, $d$ is the characteristic length of graphite particle; $\delta(d)$ is the thickness of the thin-film; $\alpha(d)$ is the dispersion coefficient of the graphite particle.

The normalized theoretical efficiency is shown in Fig 3b. The trends of the theoretical and experimental results agree well with each other, which means that the system performance can be predicted very well by the thin-film evaporation theory.

We further increase the thin-film region by using anionic surfactant to improving the dispersion of particles. Fig 3c shows that the maximum efficiency reaches up to 84%, which is

5% higher than that of without surfactant. The optimal concentration of surfactant is around 3 to 6 g/m$^2$. The efficiency begins to decrease when the concentration is higher, which may due to that the surfactant layer increases the thermal resistance between water and graphite particles.

## 4. Conclusion

In conclusion, the energy efficiency of the steam generator reaches up to 84% when using wick, expanded polyethylene, graphite particles and surfactant. Meanwhile, the cost of the proposed generator is very low compared with other high efficiency nanotechnology. The efficiencies are 56% and 79% for evaporation from wick with and without graphite particles, respectively. The graphite particles enhance the efficiency dramatically, due to the better solar absorption, heat localization and heat transfer at the evaporation regions when particles are used. The results also indicate that the thin-film evaporation resulted from graphite particles can enhance the efficiency by at least 11%. The optimal concentration of graphite particles is 60 g/m$^2$. It's also found that the smaller the particle size, the higher the efficiency, due to more thin-film regions can be created by small particles.

## 5. Acknowledgement

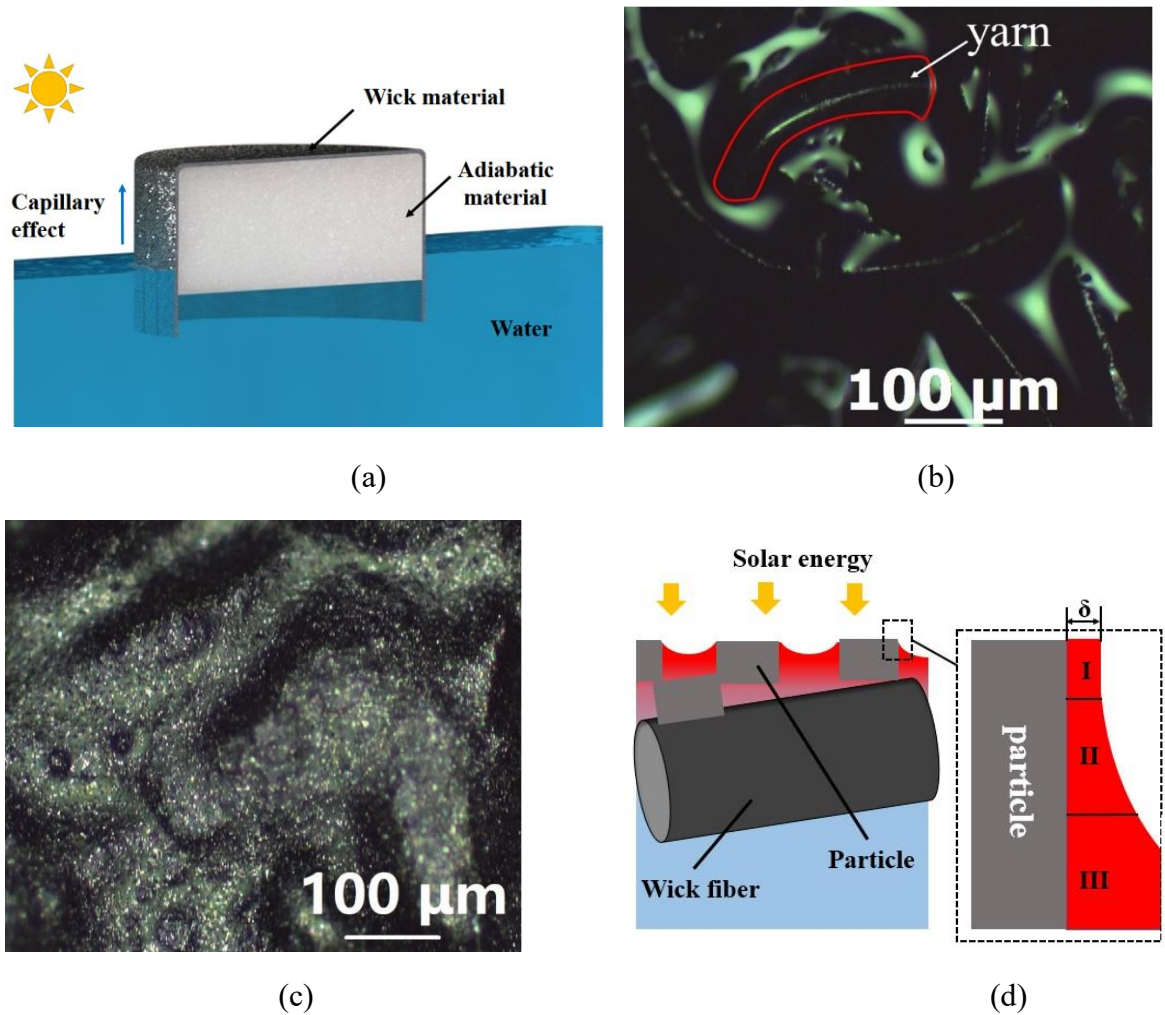

**Fig 1** a) Schematic diagram of evaporation form wick. Evaporated water is fed by capillary effect, adiabatic material is used to prevent heat loss from evaporation layer to the bulk water. b) The optical microscopic image of the wetted wick material. c) The wetted wick material covered by 60 g/m$^2$ graphite particles. the lateral size and thickness of the particles are around 1.3 μm and 100nm, respectively. d) Schematic diagram of the microscopic structure on the evaporation surface. The inset shows the meniscus near the particle. Region (I)-(III) are adsorbed region, thin film region and intrinsic meniscus region, respectively. The fast evaporation happens at thin-film region.

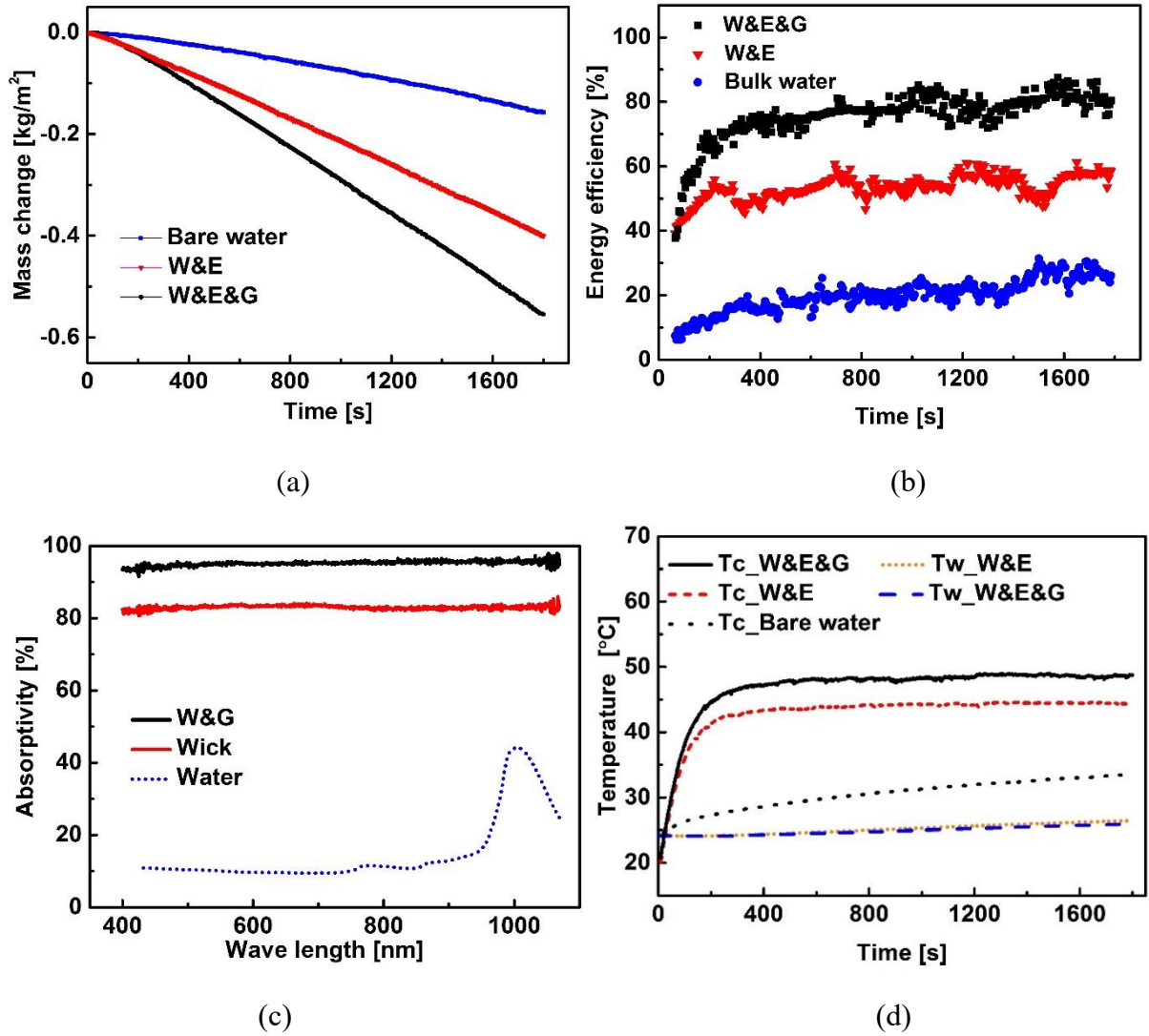

**Fig 2** a) The evaporation rates of different evaporation condition, the concentration for graphite is 60 g/m$^2$, the lateral size of particles is around 1.3 μm (10000 mesh), b) The instantaneous energy efficiency of different evaporation condition, c) The absorptivity of water, wick, wick with graphite particles (W&G). The wick material is wetted during the measurement. d) The temperature of different cases, $T_c$ is the temperature at the center of the wick layer surface or bare water surface, $T_w$ is the temperature of water 1 cm below EPE.

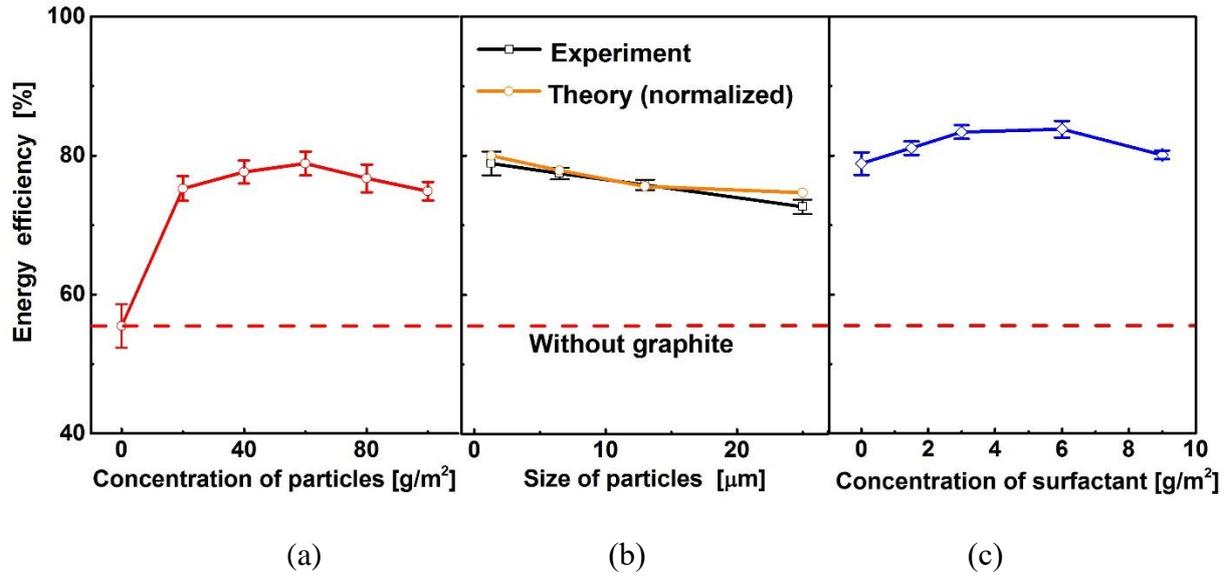

Fig 3. The effect of different parameters on the energy efficiency. The red dash line indicates the efficiency of evaporation from wick without graphite particles. a) The energy efficiency for different graphite concentration on wick material, b) The energy efficiency for different graphite size on wick material and the theoretical prediction according to the thin-film theory, c) The effect of surfactant on the efficiency.